\documentclass[fleqn,twoside]{article}
\usepackage{espcrc2}
\usepackage{epsfig}

\newcommand{\ZETA}{\mbox{\boldmath $\zeta$}}

\usepackage{graphicx}
\usepackage[figuresright]{rotating}

\newcommand{\AmS}{{\protect\the\textfont2
  A\kern-.1667em\lower.5ex\hbox{M}\kern-.125emS}}

\title{Low lying nucleons from chirally improved fermions\thanks{Presented 
at LATTICE 2003 by P.\ Crompton. 
The work was supported by the Austrian Academy of Sciences 
(APART 654, C.\ Gattringer), by Fonds zur F\"orderung der 
wissenschaftlichen Forschung in \"Osterreich (P14806-TPH, L.Ya.\ Glozman and
P16310-N08, C.B.\ Lang) and by DFG and BMBF.}}

\author{{Dirk Br{\"o}mmel$^a$, Peter Crompton$^{a}$, 
Christof Gattringer$^{a}$, 
Leonid Ya.\ Glozman$^b$, C.B.\ Lang$^b$, 
Andreas Sch{\"a}fer$^a$, Stefan Sch{\"a}fer$^a$
\hspace{1mm}(for the BGR [Bern-Graz-Regensburg] Collaboration)} 
\vskip5mm
$^a$ Institut f{\"u}r Theoretische Physik, Universit{\"a}t
Regensburg, D-93040 Regensburg, Germany. 
\vskip2mm
$^b$ Institut f{\"u}r Theoretische Physik, Universit{\"a}t
Graz, A-8010 Graz, Austria.
\vskip3mm}
       
\begin{document}
\begin{abstract}
We report on our preliminary results on the low-lying excited nucleon spectra
which we obtain through a variational-basis formed with three different
interpolators. 
\vspace{1pc}
\end{abstract}
\maketitle
There are two competing pictures for 
understanding baryon physics in different mass regions. 
For hadrons with heavy quarks  
linear confinement and perturbative gluon exchange 
corrections \cite{DeGeGl75} provide
a satisfactory description. For lighter quarks towards 
the chiral limit we have to expect
manifestations of spontaneous chiral symmetry breaking \cite{GlRi96}. 
Lattice studies can shed light on the relevant mechanisms.
In particular after the advent of Ginsparg-Wilson fermions it is now also 
possible to run computations relatively close to the chiral limit.

However, exact chiral fermion actions 
(like the overlap action) are very costly. 
On the other hand,
recently developed approximate solutions of the Ginsparg-Wilson equation 
allow to work at relatively small quark masses at only moderate numerical cost.
Two such approximate solutions, the {\sl parametrized fixed point action} 
and the {\sl chirally improved action}, have
been studied in the BGR-collaboration \cite{BGR1,BGR2}.
 Here we present results 
for low lying nucleon states computed with the chirally improved (CI) action 
\cite{chirimp}. The CI operator is a parameterization of the Dirac 
operator with terms essentially restricted to the 
hypercube. It allows for simulations
at pseudoscalar-mass to vector-mass ratios down to $m_{PS}/m_V$ = 0.32 for 
a relatively large lattice spacing of $a = 0.15$ fm                 
(see \cite{BGR1} for more details). Thus we can work with light pions on 
large physical volumes.
The quark propagators are determined for quenched gauge
configurations generated with the L\"{u}scher-Weisz action. 
Here we discuss results derived from 100 configurations on 
$16^3 \times 32$ and $12^3 \times 24$ lattices at a lattice 
spacing of $a = 0.15$ fm
as determined with the Sommer parameter \cite{GaHoSc02}. This gives rise 
to spatial extensions of 2.4 fm and 1.8 fm.
In this study our quark mass values cover the region $0.02 \leq a\,m_q\leq 0.2$
corresponding to values of $m_{PS}/m_V$ down to 0.397. A detailed account 
of our setting is given in \cite{Br0373}.

We implemented the following three interpolating operators for the nucleon
\begin{eqnarray}
\chi_1(x) & = & 
\; \;  \epsilon_{abc} \left[u_a^T(x)\,C \,\gamma_5\,d_b(x)\right] u_c(x) \; ,
\label{F1}
\\
\chi_2(x) & = &
\; \;  \epsilon_{abc} \left[u_a^T(x)\,C\, d_b(x)\right] \gamma_5 \,u_c(x) \; ,
\label{F2}
\\
\chi_3(x) & = & 
i \, \epsilon_{abc} \left[u_a^T(x)\,C\, 
\gamma_{0}\gamma_5d_b(x)\right]u_c(x) \; .
\label{F3}
\end{eqnarray}
These operators were Jacobi-smeared at both source and sink.
We remark that we also experimented with point sinks and wall sources 
but found unchanged results and no improvement of the quality of our data. 
We computed all cross correlations ($n,m = 1,2,3$)
\begin{equation} 
C_{n m} ( t )  \; = \; 
\langle {\chi_n}^{*}(0) \; \chi_m(t) \rangle \;.
\label{corrmat}
\end{equation}
For a basis of infinitely many operators ($n,m = 1,2 \, ... \, \infty$)
the diagonalization of the matrix $C$ 
would lead to the optimal operator combinations building the
physical states. Finding these combinations is equivalent to the 
generalized eigenvalue problem \cite{Lu90}
\begin{equation}
C( t ) \; {\ZETA}^{(k)}(t) \; \; = \; \; \lambda^{(k)} ( t , t_0 ) \;
C( t_0 ) \; {\ZETA}^{(k)}(t) \; ,
\label{geneval}
\end{equation}
with eigenvalues behaving as
\begin{equation}
\lambda^{(k)} ( t , t_0 ) \; \; = \; \; e^{- ( t - t_0 ) W_k}\;.
\label{expdecay}
\end{equation}
Each eigenvalue corresponds to a different energy level $W_k$ dominating its 
exponential decay. The optimal operators $\widetilde{\chi}_i$ 
which have maximal overlap
with the physical states are linear combinations
of the original operators $\chi_i$,  
\begin{equation}
\widetilde{\chi}_i \; \; = \; \; \sum_j c^{(i)}_j \, \chi_j \;  .
\label{coefficients}
\end{equation}
The coefficients $c^{(i)}_j$ are obtained from the $j$-th entry of 
the $i$-th eigenvector (corresponding to eigenvalue $\lambda^{(i)}$).
Note that here we work with only a finite basis of operators which gives rise 
to corrections to the single exponential decay Eq.\ (\ref{expdecay}) which 
are discussed in more detail in \cite{Br0373,Lu90}.

We project the operators in the correlation matrix
(\ref{corrmat}) to definite parity. 
This allows us to disentangle states with positive and negative parity.
We use the eigenvalues of the full correlation matrix to identify the plateaus 
in the effective mass. 
In this region we then use fully correlated two-parameter fits 
for the largest and second largest eigenvalues to determine the energy of the 
ground state and first excited state. We also experimented with 
Bayesian fitting techniques but do not discuss these results here.

In our data we clearly identify the nucleon, the two 
lowest negative parity states N(1535) and N(1650) and an
excited state of positive parity with a mass 
well above the masses of the two excited
negative parity states and thus too high to be identified with the
Roper state at 1440 MeV. We denote this state as $N'_+$. 
For negative parity and for
the excited state in the positive parity sector the
quality of the fits decreases towards smaller quark masses
and we cannot maintain the standards of our fitting
procedure. We do not give results in these cases.

\begin{figure}
\centerline{\epsfig{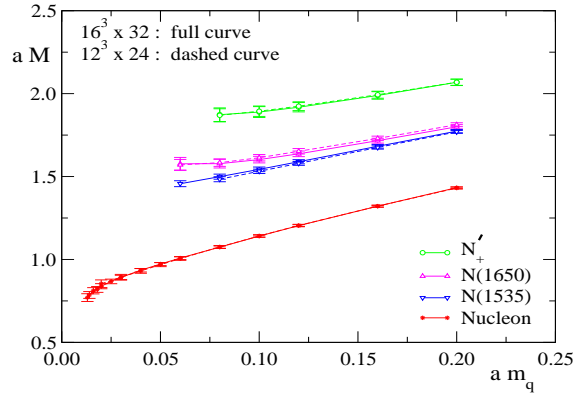}}
\vspace{-5mm} 
\caption{Results for the nucleon masses as a function of the quark mass 
(all in lattice units). We compare data from $16^3 \times 32$ and 
$12^3 \times 24$ lattices.
\label{fig_spectrum}}
\vspace{-4mm}
\end{figure}

Our results for masses the are given in Fig. \ref{fig_spectrum}. 
Like previous studies \cite{sasakietal,Me03,edwardsetal,dongetal} 
we clearly observe a splitting between the nucleon and 
the negative parity states.
At large quark masses this is attributed to the orbital excitation 
of the quark motion in the confining color-electric field. 
The splitting slowly increases towards the chiral limit. This 
implies that near the chiral limit there is another mechanism contributing
in addition to confinement. This is quite consistent with the
chiral constituent quark model \cite{GlRi96} where a considerable part of the
splitting is related to the flavor-spin residual 
interaction between valence quarks.
Also the splitting between the two negative parity states is clearly visible. 

At smaller quark masses (below $a\,m_q=0.05$) 
plateaus in the effective mass plots 
for both negative parity states and for $N'_+$ completely disappear and hence 
we cannot trace these states closer to the chiral limit. At small Euclidean
time separation there is a hint from our 
data (for smeared source and point sink) that 
there may be one more positive parity excited state.
However, our analysis did not provide consistent 
results for this weak signal.

We performed our analysis for two different lattice sizes 
$16^3 \times 32$ (full curves in Fig.\ \ref{fig_spectrum}) 
and $12^3 \times 24$ 
(dashed curves). Fig.\ \ref{fig_spectrum} shows that 
the two data sets fall on top of each other and we do not suffer from
finite size effects.

\begin{figure}[t]
\centerline{\epsfig{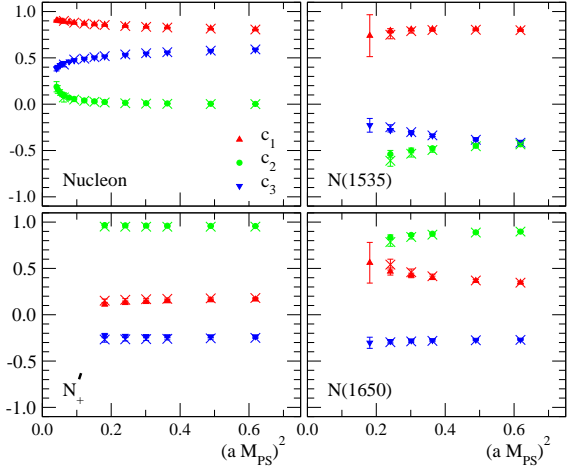}} 
\vspace{-5mm}
\caption{ 
The mixing coefficients $c_i$ of the optimal operators plotted as a 
function of pseudoscalar mass. By definition, the basis of the 
spectral decomposition has $|c_1|^2 + |c_2|^2 + |c_3|^2 = 1$. Data from 
$16^3 \times 32$ is plotted with filled symbols, for 
$12^3 \times 24$ we use crosses.
\label{comps_vs_m}}
\vspace{-4mm}
\end{figure}

Interesting physical insight can be obtained from the mixing coefficients 
$c^{(i)}_j$ of Eq.\ (\ref{coefficients}). In Fig.\ \ref{comps_vs_m} 
we show their dependence on the pseudoscalar mass and again compare 
$16^3 \times 32$ (filled symbols) and $12^3 \times 24$ (crosses) 
to illustrate the absence of finite volume effects. Operator $\chi_1$
clearly dominates the nucleon at all quark masses, 
whereas in QCD sum rule analyses
\cite{QDCsumrules} it is assumed that the equally weighted superposition of 
$\chi_1$ and $\chi_2$ is the optimal combination coupling to the nucleon.

The nucleon belongs to a 
{\bf 56}-plet which would require that under the permutation of two quarks a 
diquark subsystem has positive parity since the color wave function is 
completely antisymmetric. In the large $N_c$ limit the nucleon should not 
contain pseudoscalar or vector diquark subsystems and one expects the 
interpolator content from $\chi_2$ and $\chi_3$ to be minimal
as is indeed seen in our data.

Fig.\ \ref{comps_vs_m} shows that at large quark masses the $N(1535)$ is 
dominated by $\chi_1$ and $N(1650)$ by $\chi_2$. As the chiral limit is 
approached $N(1535)$ couples optimally to $(\chi_1 - \chi_2)$ while $N(1650)$ 
couples to $(\chi_1 + \chi_2)$ and the $\chi_3$ 
contribution is suppressed in both 
cases. $N(1535)$ and $N(1650)$ belong to the negative parity $L=1$
{\bf 70}-plet of $SU(6)$ and both contain scalar and 
pseudoscalar diquark components
in their wave functions. 
So the mixing in the vicinity of the chiral limit is 
very different from the mixing
in the heavy quark region. Within the quark model this mixing is attributed to the
tensor force and our results hint at its relation to 
spontaneous chiral symmetry breaking.

In \cite{dongetal} the subtraction of a parametrization of possible 
quenched artifacts ($\eta' N$), giving rise to negative values of the
correlation function for central $t$-values, was crucial for the
identification of a Roper signal. We also observe such negative values
but postpone a detailed analysis to future studies with better statistics.

\end{document}